\documentclass[10pt,twocolumn]{article}

\usepackage{isqcmc2023}
\usepackage{graphicx,url}

\usepackage[utf8]{inputenc}
\usepackage{url}
\usepackage{bytefield}
\usepackage{amsmath}
\usepackage{amssymb}
\usepackage{amsfonts}
\usepackage{amsthm}
\usepackage{braket}
\usepackage{verbatim}
\usepackage{tikz}
\usepackage{qcircuit}
\usepackage{algorithm2e}
\RestyleAlgo{ruled}
\usepackage{float}

\usepackage{enumitem}
\setlist[itemize]{leftmargin=15px}
\setlist[enumerate]{leftmargin=15px}
\usepackage{orcidlink}
\title{Quid Manumit - Freeing the Qubit for Art\\ \large{Embedded Quantum Simulators for Musical Instruments}}

\author{Mark Carney\inst{1}\orcidlink{0000-0001-9372-9033}}

\address{Quantum Village Inc.\\
         Delaware, USA/London, UK
         \email{mark@quantumvillage.org}
}

\begin{document}

\maketitle

\begin{abstract}
This paper describes how to `Free the Qubit' for art, by creating standalone quantum musical effects and instruments. Previously released quantum simulator code for an ARM-based Raspberry Pi Pico embedded microcontroller is utilised here, and several examples are built demonstrating different methods of utilising embedded resources: The first is a Quantum MIDI processor that generates additional notes for accompaniment and unique quantum generated instruments based on the input notes, decoded and passed through a quantum circuit in an embedded simulator. The second is a Quantum Distortion module that changes an instrument's raw sound according to a quantum circuit, which is presented in two forms; a self-contained Quantum Stylophone, and an effect module plugin called 'QubitCrusher' for the Korg Nu:Tekt NTS-1. This paper also discusses future work and directions for quantum instruments, and provides all examples as open source. This is, to the author's knowledge, the first example of embedded Quantum Simulators for Instruments of Music (another QSIM).
\end{abstract}

\section{Introduction}\label{sec:intro}

Quantum Music is a new and inspiring take on the possible futures of digitized and synthesized sound. Previous work, much of which can be found in \cite{Miranda2022}.

With the advent of free and pay-as-you-go Quantum Compute-as-a-Service (QCaaS), such as services offered by IBM and on the AWS Braket platform respectively, it has never been easier nor more accessible to perform quantum computing tasks on real quantum hardware. 

We have also seen an explosion in new simulators, possibly the most impressive being a 256-qubit quantum simulator developed by Ebadi \emph{et al.} in \cite{Ebadi2021}. 

Whilst we await the promise of the coming power offered by real quantum hardware, high-quality quantum simulators have been indispensable to both students and professionals alike in learning the inner workings of Quantum Information Theory, such as found in now classic textbooks like \cite{Nielsen2012}.

Previous work from Quantum Village has already dealt with ways in which quantum circuits can be sonified directly - producing the ability to `hear' entanglement and superposition by means of blending tones and overtones based on the statevector of a system. This project, called Soniq, is freely available FOSS, and can be found in \cite{Carney_Soniq}.

Although there have been many investigations and research projects leading to the creation of many quantum synthesizers \cite{Miranda2022}, the aim of this project was to create self-contained standalone devices that can show how to `add quantum' in ways that become natural parts of the audio processing chains in use today.

To this end, the investigations for this paper focused on creating small effects units and devices that interact with common audio communications protocols, the best example being MIDI. Along the way, a number of methods were found to embed quantum circuit simulations into existing devices - both by the physical addition of embedded quantum simulators, but also quantum-based plugins for widely available synthesizers. 

Putz and Svozil in \cite{Reck2022} consider whether quantum art will ever reveal something that classical (in many senses) art could not? Does quantum expressibility, with access to exclusively quantum notions such as entanglement and superposition, enhance the musical experience beyond that which classical or contemporary music has been able to so far? This paper does not purport to have an answer, but instead to have been driven to create tools with which musicians may begin to explore such notions. 

\section{Preliminaries}\label{sec:prelim}

We introduce the various players in the development of our quantum sonic modules; the raspberry pi pico, quantum circuits and their simulation, digital signal processing, and MIDI.

\subsection{Raspberry Pi Pico}

Released in January 2021 \cite{raspberrypi}, the Raspberry Pi RP2040, a custom-made ARM-based microcontroller has fast become a tool of choice for many and varied applications. 

The features specifically useful to our purposes include its wide availability, low price, powerful ARM Cortex-M0+ dual core processor running at 133MHz, with 264KB RAM and many peripherals, including 8 programmable PIO pins, 30 GPIOs and built in integrator and integer divider peripherals. 

Many of these features will be used in these projects, especially the ability to run two separate fast ARM cores separately - one of which will handle signals acquisition and processing, whilst the other runs our bespoke quantum simulator code.

\subsection{The Embedded Quantum Simulator} 

\begin{figure}[t]
\centering
  \includegraphics[width=.45\textwidth]{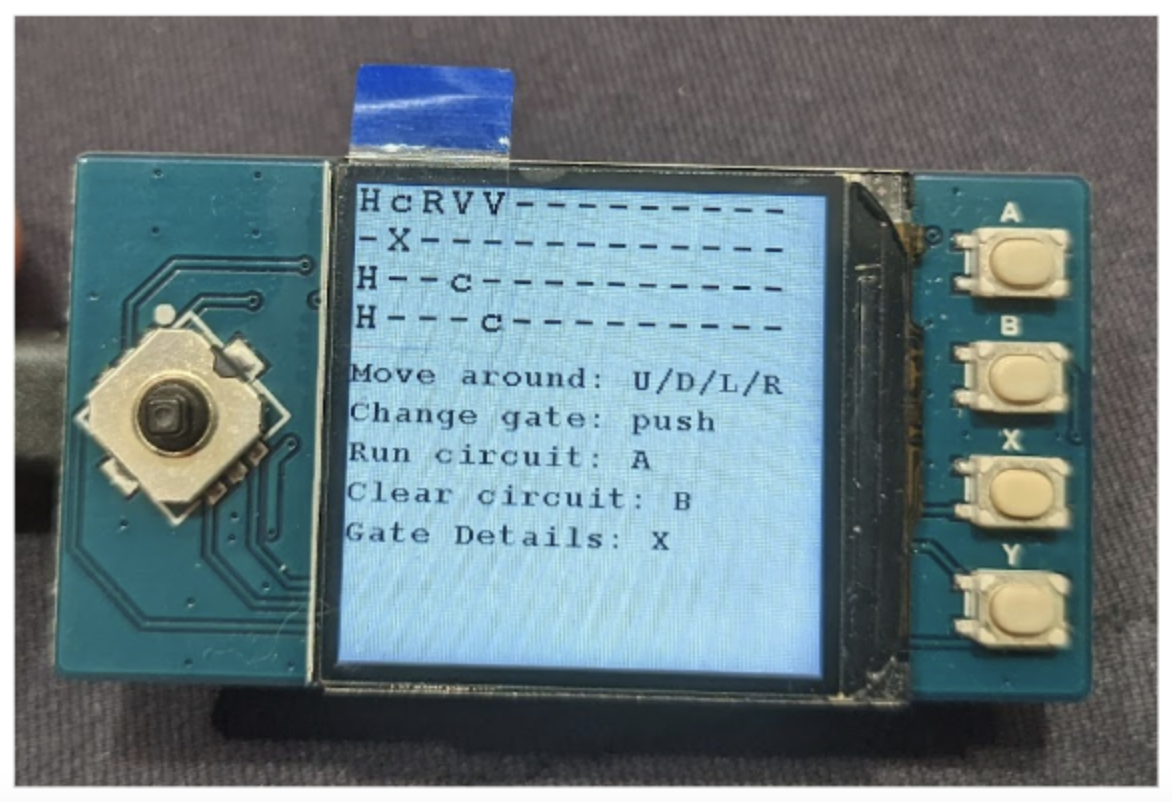}
\caption{An assembled and running microcontroller based quantum simulator, here showing a CHSH game quantum circuit.}
\label{fig:microqsim}
\end{figure}

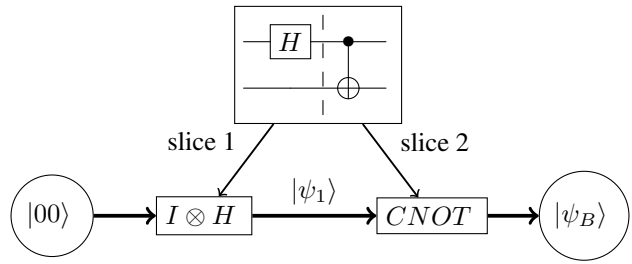
\begin{figure*}[t]
\begin{center}
\begin{tikzpicture}[every text node part/.style={align=center}]
\node (a) [circle, draw] at (0,0) { $\ket{00}$ };
\node (b) [rectangle, draw] at (2,0) { $ I \otimes H $ };
\node (c) [rectangle, draw] at (5,0) { $CNOT$ };
\node (1) [rectangle, draw] at (3.5,2) { $ \Qcircuit @C=1em @R=.7em {& \gate{H} \barrier{1} & \ctrl{1} & \qw \\ & \qw & \targ & \qw} $ };
\node (d) [circle, draw] at (7,0) { $\ket{\psi_{B}}$ };
 
\draw [->, line width=0.5mm] (a) -- (b);
\draw [->, line width=0.5mm] (b) -- node [above] {$\ket{\psi_1}$} (c);
\draw [->, line width=0.5mm] (c) -- (d);
\draw [->, line width=0.25mm] (1) -- node [above left] {slice 1} (b);
\draw [->, line width=0.25mm] (1) -- node [above right] {slice 2} (c);
\end{tikzpicture}
\end{center}
\caption{A diagram showing how circuits are split into slices, which are subsequently processed in the embedded simulator. The output $\ket{\psi_{B}} = \frac{\ket{00} + \ket{11}}{\sqrt{2}}$ is achieved by passing the initial statevector through the $I \otimes H$ tensor product matrix, and then the output $\ket{\psi_1} = \frac{\ket{00} + \ket{01}}{\sqrt{2}}$ through a $CNOT$ matrix.}
\label{fig:simprocess}
\end{figure*}

\subsubsection{Simulator Overview}

We utilised an embedded quantum simulator that the author had previously released in \cite{Carney_Micro_Quantum_Simulator_2022} for the first Quantum Village at the annual DEF CON hacking conference in Las Vegas, Aug 2022. Figure \ref{fig:microqsim} shows the original assembled module that was available at the conference. This simulator was designed as a very lightweight statevector simulator, with a low memory footprint but still calculating complex-valued statevectors from given quantum circuits.

Circuits are programmed by taking 4-qubit circuits up to 14 `slices' long - effectively treating a circuit as a column of $2 \times 2$ complex matrices. Pre-programmed gates are assigned to each position on each slice.

The simulator takes each slice of the circuit and calculates the tensor product. On the fly, it makes adjustments for control-gates allowing them to be added arbitrarily and the operator matrix is calculated for whichever gate is being controlled in any position. Figure \ref{fig:simprocess} shows an example of how this works. 

\subsubsection{Simulator Operation}

A starting statevector of $\ket{0000}$ is taken, and multiplied by each calculated slice matrix successively - the output statevector from the previous slice becoming the input for the next. This method means that we only need to have up to 3 full-size operator matrices in memory at any one time - one for the current slice, and 2 for handling various gates (e.g. CNOT and SWAP gates in particular). The final statevector is then correct for the given circuit following this composition operation. An outline of this process can be found for a 2-qubit entanglement circuit in figure \ref{fig:simprocess}.

Although the simulator was originally written for handling 4-qubit circuits across 14 slices, we only use 2 qubit circuits with length of 6 slices in these modules. This is mostly to make sure that the simulator is operating as quickly as possible. It can, of course, be extended.

\subsubsection{Interface and Hardware}

A Waveshare Protoyping board was used to build the interface as it comes complete with 2-axis joystick, 5 switch buttons, and a small LED screen. A very basic text-based interface was designed to allow the player to input gates from a fixed set of options. 

This interface takes up a lot of compute time on the RP2040, and so was omitted for these prototypes. However, with the provision of dual cores, it might be possible to offload the slower operations for the screen to the core already doing slower work for the simulator, especially in the case of the Quantum MIDI module we shall describe below.

\begin{figure*}[!htb]
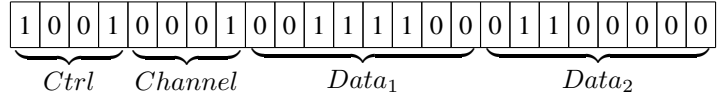

\begin{center}
\begin{bytefield}[bitwidth=1.1em]{21}
\bitbox{1}{1} & \bitbox{1}{0} & \bitbox{1}{0} & \bitbox{1}{1} & \bitbox{1}{0} &
\bitbox{1}{0} & \bitbox{1}{0} & \bitbox{1}{1} & \bitbox{1}{0} & \bitbox{1}{0} &
\bitbox{1}{1} & \bitbox{1}{1} & \bitbox{1}{1} & \bitbox{1}{1} & \bitbox{1}{0} &
\bitbox{1}{0} & \bitbox{1}{0} & \bitbox{1}{1} & \bitbox{1}{1} & \bitbox{1}{0} & \bitbox{1}{0} & \bitbox{1}{0} & \bitbox{1}{0} & \bitbox{1}{0} \\

\bitbox[]{4}{$\underbrace{\hspace{4em}}_{\text{\normalsize $Ctrl$}}$} &
\bitbox[]{4}{$\underbrace{\hspace{4em}}_{\text{\normalsize $Channel$}}$} &
\bitbox[]{8}{$\underbrace{\hspace{8.3em}}_{\text{\normalsize $Data_1$}}$} &
\bitbox[]{8}{$\underbrace{\hspace{8.3em}}_{\text{\normalsize $Data_2$}}$} \\
\end{bytefield}
\end{center}
\caption{The structure of a MIDI message. \\ This message indicates a device to start playing a C4 note on channel 1.}
\label{fig:midimessages}
\end{figure*}

\subsubsection{Simulated Measurements}

\begin{algorithm}[t]
\KwIn{A statevector $\ket{\psi}=\langle a_1, a_2, a_4, a_4\rangle$ of complex values}
\KwOut{Two random bits from the distribution defined by the statevector}
$rand \gets \text{random value, } 1 \leq rand \leq 100 $\;
$D \gets \{d_1,d_2,d_3,\ldots,d_{100}\}$\;
$B = \{b_1, b_2, b_3, b_4\}$\;

\For{$i \gets 1$ \text{ to } $4$}{
    $b_i = \lfloor 100 \times |a_i |^2 \rfloor$ \;
}
$p \gets 0$\;
\For{$j \gets 1$ \text{ to } $4$}{
    \For{$k \gets 1$ \text{ to } $b_j$}{
        $d_p = (j - 1)$\;
        $++p$\;
    }
}
$r \gets d_{rand}$\;
\Return{$r$}\;
\caption{Measurement Simulator}
\label{algo:meas}
\end{algorithm}

The simulator is also capable of generating actual simulated measurements from a circuit. It does this by taking the squared absolute values of each element in the statevector. These values are then used to populate a distribution from which a value is picked at random by the RP2040's built in randomness ring oscillator - a ring of an odd number of NOT gates that acts as a source of true randomness for the microcontroller. 

The algorithm used to generate and select from the distribution given by a state vector is found in Algorithm \ref{algo:meas}. 

This method is based on the fact that statevectors are essentially time-independent abstract probabilistic distributions of states. Given, for a random variable $X$ and statevector $\ket{A} = \{a_1, a_2, \ldots a_n\}$: $$ P(X \in [1,2,\ldots,m]) = \sum_{i=i}^{m}|a_i|^2 $$ which follows easily from Born rule: $$ \sum_{i=1}^{n} |a_i|^2 = 1 $$ The method presented above is therefore a reasonable approximation of the actual behaviour of the quantum system we are simulating.

Thus if we calculate the distribution and place this into an array, a truly random selection from this distribution array models an effective measurement of a quantum state. In this sense, our module is closest to being quantum given a high-quality random source.

\subsection{Digital Signal Processing}

Amplifiers of various kinds have been the driving force for integrated circuits and microchips since their invention (cf. Fairchild's early work in \cite{Lojek}). However, in the 80's much of this became digital processing, and the Digital Audio Workstation (DAW) was born \cite{kefauver2007}.

Indeed, as microprocessors have become faster, more efficient, and with more memory and resources built into single chip designs it has become easier than ever to do audio processing for very minimal and cheap resources compared to early DAW requirements.

The chosen RP2040 has several built in Analogue-Digital Converter (ADC) peripherals that can sample at full clock speed. There are also provisions in firmware for generating `analogue' output by means of a Pulse-Width Modulated (PWM) signals from any output pin. A PWM signal can be smoothed with an RC network to produce an analogue level. 

The only additions to our microcontroller board will be these output smoothing networks, as well as some Operational Amplifiers (opamps) for handling input/output levels from instruments and to other equipment. 

\subsection{MIDI Overview}\label{sec:midi}

The launch of MIDI in 1987 changed the way we `do' music forever. It enabled the considered automation of nearly every part of a musical experience, through a very robust and scalable communications infrastructure \cite{Huber1991-zf}. Whilst many other protocols from the same era have since become packet based and modernized (e.g. CANbus becoming CAN Ethernet \cite{canether}), MIDI has persisted in studios and DAWs worldwide. 

At its heart, MIDI is an essentially basic and simple protocol based on UART serial communications over a baud rate of 31.25kHz. It has a very simple structure for each message - each is 2 or 3 bytes, comprising the structure depicted int figure \ref{fig:midimessages}. This message shows a `note on' (\verb|0x90|) on channel 1 (\verb|0x01|) with note C4 (\verb|0x3C|) with velocity 96 (\verb|0x60|).

Given the simplicity of these messages, we can easily create code to interact with MIDI devices by reading/writing the appropriate messages using a microcontroller, and then feeding these messages back with additional messages generated by our Quantum Simulator.

\section{Module Architecture}

For both of these modules we use the same overall approach to creating the quantum effects processing.

\subsection{Sonification Methods}

Owing to the very short time-frames within which we have to both perform basic digital signal processing \emph{and} provide the results of a quantum simulation, the approach undertaken in this work is very minimal. 

Predominantly the strategy is based around basic additive/subtractive synthesis of sound. For the MIDI module the quantum simulator provides an `adjustment' added on top of the player's note. For our distortion effect module the quantum simulation result is added or subtracted from the value taken from the ADC itself before it is sent to the PWM output. 

Such meagre approaches need not be limiting, especially as the output of the simulator is based upon the player's input in both cases. We are hearing the quantum circuit respond to the input from the player, who will surely respond back completing a feedback loop with some artistic promise. 

\subsection{Quantum Circuit}\label{sec:qcirc}

This section uses standard notation, a reference for which is \cite{Nielsen2012}. The following parameterized entangler circuit is hard-coded into the firmware:

$$ \Qcircuit @C=1em @R=.7em {& \lstick{\ket{0}} & \gate{H} & \ctrl{1} & \meter & \qw \\ & \lstick{\ket{0}} & \gate{R_x(\pi s)} & \targ & \meter & \qw} $$
Here, $s$ denotes a normalised input from the player. 

This choice of circuit is based on the ability of the lower qubit to separate the two entangled state possibilities. Let $$ \ket{\Phi^+} = \frac{\ket{00} + \ket{11}}{\sqrt{2}}, \ket{\Psi^+} = \frac{\ket{01}+\ket{10}}{\sqrt{2}} $$ such as is standard notation. Note the identity: $$ \Qcircuit @C=1em @R=.7em {& \gate{R_x(\theta)} & \qw} = \Qcircuit @C=1em @R=.7em {& \gate{ph(-\theta/2)} & \gate{X^{\theta/\pi}} & \qw} $$ where $\Qcircuit @C=1em @R=.7em {& \gate{ph(\alpha)} & \qw } = e^{i\alpha} \mathbb{I}$. 

We may disregard the global phase shift induced by $R_x$ and then see that for $0\leq s \leq 1$ our state may be represented by: $$\frac{(\cos(s/2) \times (\ket{00}+\ket{11})) + (\sin(s/2) \times (\ket{01}+\ket{10}))}{\sqrt{2}}$$ when ignoring global phase. This can be further refined as a combination of the Bell states above: \begin{align}\label{eqn:state} \big( \cos(s/2)\ket{\Phi^+} + \sin(s/2)\ket{\Psi^+} \big) \end{align} 

By applying the $R_x$ gate to the lower qubit, we can use it to `shift' between the two entangled states and even obtain a balanced superposition of these states, \emph{i.e.} total disentanglement at a quarter-turn value of $s$. So this circuit allows us to entangle, disentangle, and re-entangle differently, through a single parameter as we please.

For the MIDI example, the note from the player is taken and divided by 63 (which is $\lfloor{\frac{127}{2}}\rfloor$) to ensure that $0 \leq s \leq 2$. This is so that the note fits within the $2 \pi$ rotations possible about the $R_x$ rotation. The way that the input note will affect the output can be crafted to affect the registers of the notes that are given by the quantum simulator. Lower or higher notes will place the state in (\ref{eqn:state}) closer to the state $\ket{\Phi^+}$ whilst notes in the mid-range will place it closer to the state $\ket{\Psi^+}$. 

For our purposes we place the output bit pairs from the quantum simulator sequentially into an 8-bit output. With $\ket{\Psi^+}$ this has a higher likelihood of generating higher values, owing to the higher probability of `11' appearing in the MSBs of the output integer. The converse is true for $\ket{\Psi^+}$, which is more likely to give closer values as 1's are less likely to be in more significant bits. 

For the distortion we allow any value to be input, which is fine as we can rely on the cyclic nature of $\cos$ and $\sin$ to ensure our input is still valid, controlled in both examples below by a physical user interface, usually a potentiometer.

With these rotations set in code, the quantum circuit code can be fixed to be as efficient as possible, greatly reducing the complexity and time taken to compute.

\subsection{Software Structure}

In both modules we take the same approach:
\begin{itemize}
    \item Separate out the music processing from the quantum processing.
    \item Utilise the build in Interrupt Requests (IRQs) and interrupts to drive activity based on player input, minimising any latency or lag.
    \item Optimize the quantum circuit to run in as short a time as possible.
    \item Choose an appropriate number of samples for the measurement distribution. 
    \item Pass the output from the quantum simulator back out to the output in some appropriate manner - the `sonification' of our quantum output.
\end{itemize}

For MIDI, we use the quantum simulator to create values in a given range, which are then added to the player's note value to create a `quantum accompaniment'. For the distortion module the quantum simulator output values are added/subtracted from the original signal to produce a distorted output, hence the name. 

All code was written in C using VSCode\footnote{\url{https://code.visualstudio.com/}}, and compiled using the \verb|pico-sdk| package found in \cite{picosdk}. Code was compiled with \verb|cmake| on both MacOS and Linux operating systems. 

The code from this project is hosted on our Github, located in \cite{Carney_QMusic}. 

\subsection{Hardware}

Both modules were prototyped on breadboards with off the shelf components and generic RP2040 boards - the Raspberry Pi Pico boards produced by Raspberry Pi and PiHut. Generic resistors and capacitors were used. MCP6004 OpAmp IC's were used for the quantum distortion module. A Hobbytronics MIDI Breakout Board was used for the quantum MIDI module. 

\section{Methodologies and Setup}

\begin{figure*}[ht]
\centering
  \includegraphics[width=\textwidth]{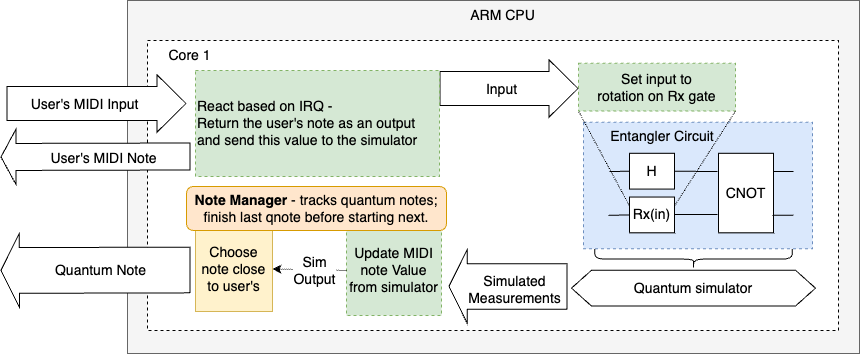}
\caption{High-level overview of the Quantum MIDI module architecture.}
\label{fig:qmidiArch}
\end{figure*}

\begin{figure}[t]
\centering
  \includegraphics[width=0.45\textwidth]{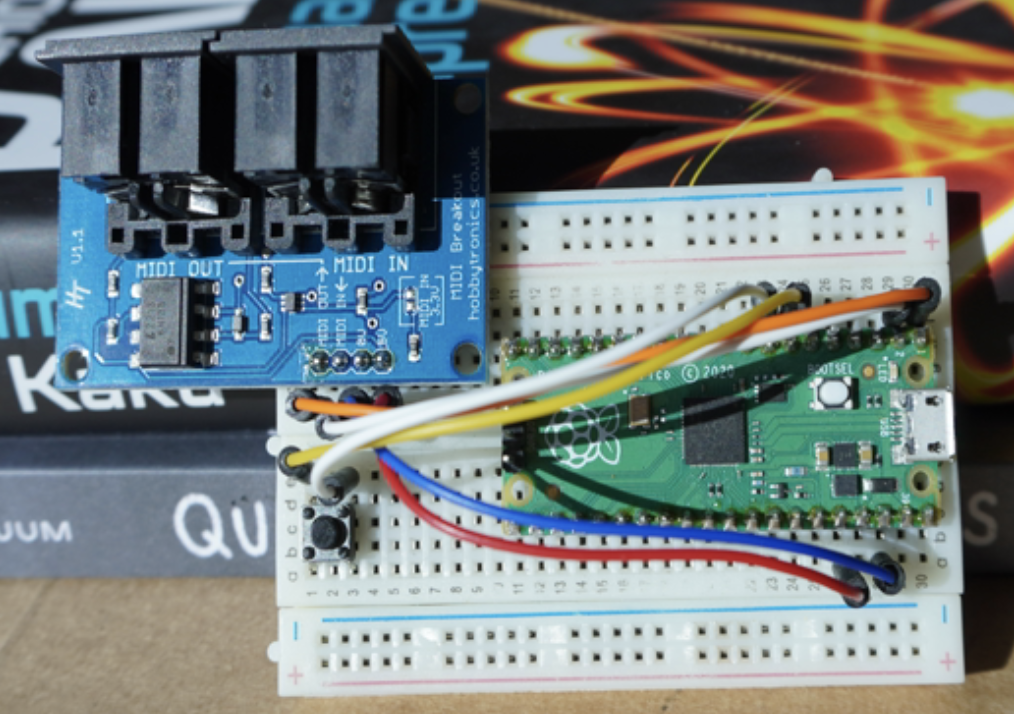}
\caption{Photograph of the breadboard prototype of the Quantum MIDI Module.}
\label{fig:qmidiPhoto}
\end{figure}

\subsection{Quantum MIDI Module Design}

This section gives details on the quantum MIDI module that was constructed. The overall architecture is given in figure \ref{fig:qmidiArch}. A photograph of hardware used to prototype the module may be found in figure \ref{fig:qmidiPhoto}.

\subsubsection{MIDI Processing}

The best way to construct a usable MIDI interface is to use interrupt-driven UART peripherals, such as the RP2040 possesses. 

Firstly, we created a routine to read any MIDI message arriving from the MIDI\_IN port and simply pipe it straight out to the MIDI\_OUT port, driven by interrupts. This was as quick as you would expect for a minimal amount of code, and any latency was negligible from a performance perspective.

\subsubsection{Adding the Quantum Simulator}

With this working, the quantum simulator was added. The simulation of a very small circuit such as that described in figure \ref{fig:qmidiArch} (in the blue box) was sufficiently fast that it can be run constantly alongside the interrupt-driven MIDI UART interface. 

The simulator would run continually in a simulate/delay loop until an interrupt was given to process a MIDI message.

If the message is not a `note-on' message (\verb|0x90|) then we capture the whole message and relay it to the output first. Then we stop any previous `q-notes' that have been started and instead send a new q-note formed from adding the note the player just gave us plus some shift given by the quantum simulator. We then pass the player's note back into the quantum simulator to get the next shift value for the next q-note. 

Given MIDI requires notes to be explicitly turned off, a timing mechanism was added to turn off any q-note after around 2 seconds wait. This helps better manage the soundscape coming from the module without it becoming too cluttered when using instruments with very slow or zero decay envelopes, such as pipe organs.

\subsubsection{Quantum Patch Generation}

To amplify the possibilities, the MIDI control codes, often called `SysEx' (System Exclusive) commands, for a Yamaha Reface DX were downloaded and coded into the Quantum MIDI controller. A button was added to control this new code path and the note-inputs into the quantum circuit were then used to generate random patch settings for the 4 oscillators. 

Upon experimentation, minimal changes were made to the main (first) oscillator for each default patch, with substantial changes being made to the subsequent oscillators with random values generated from the above quantum circuit. 

The reason for leaving the main oscillator untouched is that there is a high likelihood for making an inaudible patch if you do manipulate it too far, so by keeping this setting and only changing the others it usually maintains some reliability of the new patch being audible. 

These new code paths enables a performer to generate on the fly a totally new and different instrument, influenced by the notes they have been playing with a short memory buffer. 

\subsection{Quantum Distortion Module Design}

\begin{figure*}[ht]
\centering
  \includegraphics[width=\textwidth]{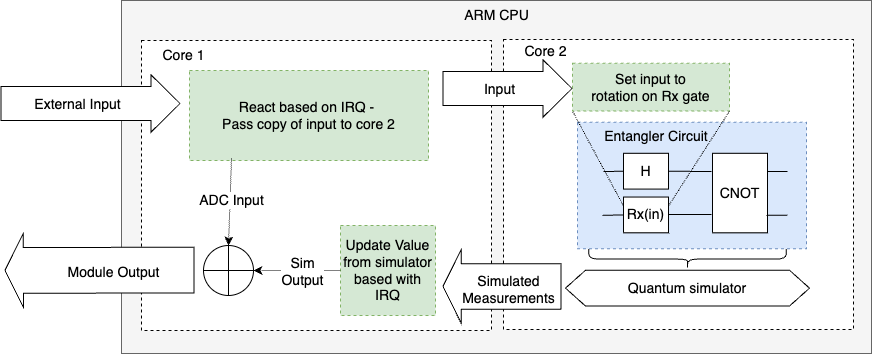}
\caption{High-level overview of the Quantum Distortion module architecture.}
\label{fig:qdistArch}
\end{figure*}

This section gives details on the quantum distortion module that was constructed. The overall architecture is given in figure \ref{fig:qdistArch}.

\subsubsection{DSP on the RP2040}

There are just 4 ADC's on the RP2040 silicon, but almost any output GPIO pin can operate as a PWM output. The former is very easy; the chosen ADC pin is setup to receive an analogue signal, and it will continually update with the value from that pin with each clock cycle. 

PWM is handled a little differently; an interrupt is setup for the `loop-back' at the end of each PWM pulse width time. We set the PWM to have the fastest resolution to try and capture as much from the player as possible. 

\subsubsection{Benefits of Dual Core}

With so much timing accuracy needed to bridge an ADC to a PWM, many options were explored to mitigate lag or latency in adding quantum. 

The Programmable Input/Output (PIO) pins - essentially small state machines that can be user programmed to take very small but repetitive computational load off the CPU cores - were found to not be particularly useful for this task. Whilst they can perform many tasks, they have no working memory and so cannot be used in the simulator. They also cannot read directly from memory, so could not be used to read the ADC value directly. 

It very quickly became clear that the best way to achieve the goals of this module - to distort sound with the output of a quantum simulator, live - we would need to utilise the dual cores of the RP2040. 

For the input to the circuit, it was also decided that the best way would be to let the user control this through a potentiometer tied to the ADC pin on the RP2040. This is then read in as the value for the circuit $R_x$ rotation. The potentiometer is mapped from 0-3.3V as ranging from 0 to 1 for values of $s$.

\subsubsection{Adding Quantum Outputs from RP2040}

The architecture in figure \ref{fig:qdistArch} shows the underlying design decisions taken to achieve very fast processing times for the quantum simulator. 

The second core was assigned to be the quantum simulator, receiving messages from the first core over an inter-core communications buffer. This buffer is driven by IRQs for both send and receive. The second core receives input from the first core, and to use this as input for a simulation each time, passing the results back as they are generated. 

The results are then combined with the current value from the ADC to form the PWM duty cycle time at that moment. This is how our quantum distortion was achieved.

\section{Results}

The modules have been effectively run to create new performances based on these strange new quantum capabilities. 

\subsection{Q-MIDI Mania}

This is an analysis the two use cases for the Quantum MIDI Module.

\subsubsection{Quantum Accompaniment}

It is very effective, although the simulation time means it reacts in a delayed manner. Because it passes-through all messages that are nothing to do with notes, it does not impede any other performance messages being sent over MIDI. Also, because it only looks for the player's notes on channel 0, you can bypass it by effective channel management. 

However, there are some bugs. For example, the use of pitch bend controls causes some very unusual and unintended sounds to be generated by the device - although these are completely unintentional, they have been found to be very pleasing. The glissandi spread over several octaves, presumably owing to the fast messaging out-pacing the interrupt management code. 

An example improvisation with the Quantum MIDI module can be heard on SoundCloud here: \url{https://on.soundcloud.com/xiATB}

\subsubsection{Quantum MIDI Patch Generation}

This patch generation tool is very effective, and generates nuanced and distinctive synthesizer patches for the Yamaha Reface DX that it was written to control. 

The main issue is that there could be more evidence for how the patches change in relation to the player's input, but this is now ongoing research and development.

\subsection{Quantum Distortion}

Although the sound generated by the Quantum Distortion module is quite jarring at times, it is given to being very distinctive. It would be nice to tune the various parameters in the code to bring out the player's note-wise input more, but the overall effect is a very gruff and impactful distortion module.

In all cases, there are two controls defined; gain, and rotation angle (denoted `$\theta$'). For the circuit outlined in section \ref{sec:qcirc}, this is all the control that we really need to edit the properties of the circuit. 

\subsubsection{Physical Implementation in a Stylophone}

To demonstrate the quantum distortion effect, a Raspberry Pi Pico was embedded within a standard Stylophone unit. The output from the simulation was sent via a parasitic added line to the TRIGGER pin on the main N555 timer within the synthesizer.

Two controls were also added - one for the gain of the extra signal to the TRIGGER pin, and one read as an analogue input by the RP2040 to control the $Rx(\theta)$ value. 

The full assembly can be found in figure \ref{fig:qStylophone}, and code and diagrams can be found in \cite{Carney_QMusic}. A sample of the sound created from this new device can be found on SoundCloud: \url{https://on.soundcloud.com/GBh9Y}

\begin{figure}[t]
\centering
  \includegraphics[width=0.45\textwidth]{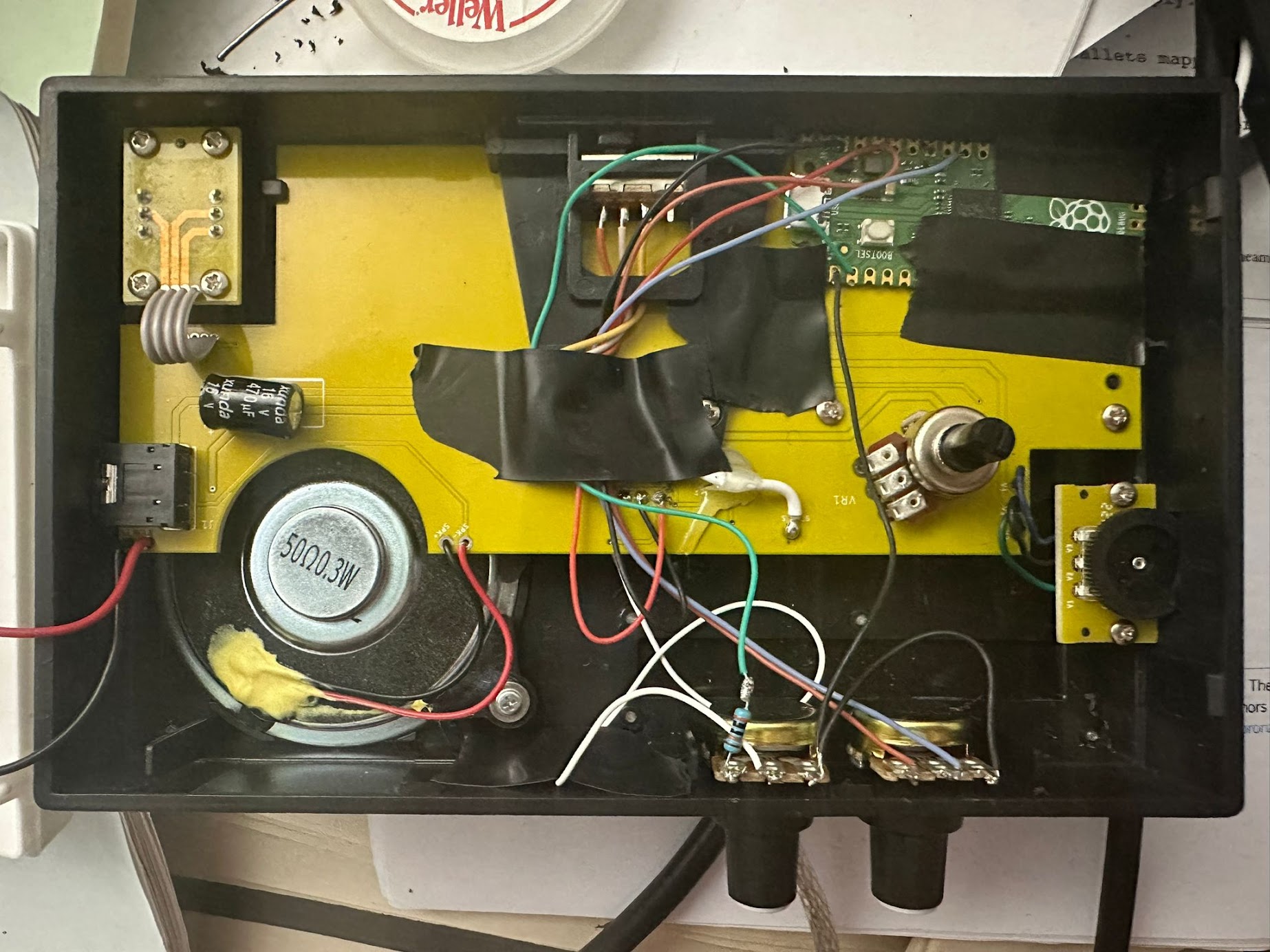}
\caption{Photograph of a Stylophone with added controls and embedded quantum simulator.}
\label{fig:qStylophone}
\end{figure}

\subsubsection{Code Implementation in a KORG Nu:Tekt NTS-1}

The code was miniturized and distilled further in order to fit into a KORG \verb|logue-sdk| module called `QubitCrusher' that could be compiled and loaded into a KORG Nu:Tekt NTS-1 synthesizer. 

The effects module is based on the `bitcrusher' distortion unit, replacing the user-defined cutoff parameters for the distortion with quantumly-generated values from the quantum circuit in section \ref{sec:qcirc}.

These relatively inexpensive units are readily available, and the same circuit used on the RP2040 hardware was repurposed as a statevector with a dramatically reduced code base. 

The code was significantly rewritten to only have the main features of the simulator: a pre-computed statevector, that is manipulated to give the outputs of the correct $\sin$ and $\cos$ values. These are then committed to the effect memory, and the simulator code from Algorithm \ref{algo:meas} is put in place to create the effect in the effect chain. 

Even without a dual core, the effect is still fast enough to work live on the KORG synthesizer, which is a fantastic result. It also has a sonic similarity of timbre and general effect comparable to the hardware based Quantum Stylophone, which is reassuring that we are hearing an identifiable phenomenon. 

This module was released free and open source on our github \cite{Carney_QMusic}, and a sample of the sounds created can be found on SoundCloud: \url{https://on.soundcloud.com/UgXe4}

\section{Conclusions}

The work presented in this paper has shown that embedded quantum simulators can indeed be used for musical purposes and the sonification of quantum states. Whilst the resources used here were quite limited, it shows how much can be done by careful consideration of hardware choices and software design architecture.

The code in this paper demonstrates the following sonification methods with embedded quantum simulator systems:
\begin{enumerate}
    \item Quantum Effects Units - using the quantum distortion code in several settings.
    \item Quantum Note Generation - using our Quantum MIDI module.
    \item Quantum Instrument Generation - using the same Quantum MIDI hardware.
\end{enumerate}

Of course, there are other solutions that fit into similarly-sized packages, notably the Pi Zero\footnote{\url{https://www.raspberrypi.com/products/raspberry-pi-zero/}} which is a full-stack computer on a small format PCB. This device almost certainly has the considerable resources to do more than our comparatively meagre microcontroller. 

The exercise carried out in this paper shows how few resources are really required to do `edge quantum computing' in a usable and efficient manner. The aim has been to demonstrate that whilst `quantum computing powered by a potato'\footnote{To be singular, the author notes that this would be a very large potato.} is hard, it is not impossible. 

\section{Future Work}

There are many directions that this work could go in, some of which are listed here. 

\subsection{WiFi to Cloud Quantum Compute}

It would be very straightforward to use a Pi Pico W module or some other WiFi-enabled microcontroller to connect to the internet over WiFi and access quantum compute resources via the cloud. 

We have written a PoC of this, but owing to queue lengths and speed issues this was not found to be an immediately useful method. If one could secure guaranteed short-notice access to a quantum computer, then this method would be very workable.

\subsection{Multi-Instrument MIDI}

With so many instruments across 16 channels, future work is currently planned to explore how multi-instrument quantum midi can be achieved. It would be interesting to see how quantum chords, and even quantum harmony can be derived from the simulation results.

\subsection{Eurorack and Pedal Packages}

The author is currently working on Eurorack and guitar pedal packages for these modules that would allow these ideas to go from breadboard prototypes to possibly workable instrument modules that would allow performers and composers to experiment with quantum in a familiar and accessible way.

\subsection{More Qubits}

Whilst the simulator is currently only using 2 qubits, it would be nice to add more. This could be done by only simulating a given statevector, constructed whenever the circuit is manipulated.

Additionally, there is work undergoing to increase the number of qubits possible in the original micro-qsim project. Currently, 6 qubits seems to be the limit, but by moving the model away from statevectors and towards tensor network simulations, we expect to increase this. This would also likely lead to speed improvements in these projects presented in this paper.

\subsection{Quantum Networked Instruments and Non-Local Ensembles}

``The quantum internet is the holy grail of quantum information processing." \cite{Azuma2022nuy} Perhaps with the addition of quantum networking between instruments we can give rise to quantum ensembles; multiple musicians and artists manipulating a communal quantum state from which new horizons of sound and music may be found. It would be quite an achievement to hear a `\emph{non-local ensemble}' perform, enabled by the addition of quantum communications to their instruments. Whether this would work best for string quartets, rock bands, or jazz trios would be an exciting exploration in musical composition.

\section*{Acknowledgements}

The author wishes to acknowledge the helpful comments and feedback from the following people: Victoria Kumaran, and Dr. Frey Wilson. This paper is also to appear at the ISQCMC conference in Berlin, Oct 5-6 2023.

\bibliographystyle{unsrt}
\bibliography{bib}

\end{document}